\documentclass[aps,prc,preprint,showpacs,showkeys]{revtex4}
\usepackage{graphicx}
\usepackage{bm}
\usepackage{epsf}

\begin{document}

\title{Modification of nucleon properties in nuclear matter and finite
nuclei }
\author{W. Wen}
\affiliation{Department of Physics, Nankai University, Tianjin 300071, China}
\author{H. Shen}
\email{songtc@nankai.edu.cn}
\affiliation{Department of Physics, Nankai University, Tianjin 300071, China}

\begin{abstract}
We present a model for the description of nuclear matter and finite
nuclei, and at the same time, for the study of medium modifications
of nucleon properties. The nucleons are described as nontopological
solitons which interact through the self-consistent exchange of
scalar and vector mesons. The model explicitly incorporates quark
degrees of freedom into nuclear many-body systems and provides
satisfactory results on the nuclear properties. The present model
predicts a significant increase of the nucleon radius at normal
nuclear matter density. It is very interesting to see the nucleon
properties change from the nuclear surface to the nuclear interior.
\end{abstract}

\maketitle





\section{Introduction}

\label{introduction}

One of the most interesting topics in nuclear physics is to study how the
nucleon properties change in nuclear medium. So far, there are many
experimental evidences indicating that the properties of the nucleon bound
in nuclei are significantly modified from those of a free nucleon. The
famous European Muon Collaboration (EMC) effect shows that the nucleon
structure functions in nuclei deviate from those in a free nucleon~\cite%
{emc1}. Important evidences for medium modifications also come from recent
polarization transfer experiments at the Thomas Jefferson National
Accelerator Facility, which observed a difference in the electromagnetic
form factors of a proton bound in a helium nucleus compared to a free one~%
\cite{emc2}. On the other hand, there are numerous theoretical works on the
study of in-medium nucleon properties based on various models~\cite%
{prc85,qmf98,npa00,ppnp07}. At present, we are still far away from
describing nucleons and nuclei in terms of quarks and gluons using quantum
chromodynamics (QCD), which is believed to be the fundamental theory of
strong interactions. Hence it is highly desirable to build models which
could incorporate quark degrees of freedom and respect the established
theories based on hadronic degrees of freedom.

The quark-meson coupling (QMC) model proposed by Guichon~\cite{qmc88} can be
considered as an extension of the extremely successful theoretical treatment
of nuclear many-body systems, known as quantum hadrodynamics~\cite{WS86}, to
include the internal structure of the nucleon. The QMC model describes the
nuclear system as nonoverlapping MIT bags in which the confined quarks
interact through the self-consistent exchange of scalar and vector mesons in
the mean-field approximation. In the QMC model, the quark structure of the
nucleon plays a crucial role in the description of nuclear matter and finite
nuclei~\cite{ppnp07}. It is also possible to study the medium
modification of nucleon properties in the QMC model where the quark degrees
of freedom are incorporated explicitly in the nuclear many-body
system~\cite{qmc95}. In the past decade, the QMC model has been extensively
developed and applied with reasonable success to various nuclear
phenomena~\cite{ppnp07,qmc96,qmccm1,qmccm2,qmc97,qmc98,qmc99,qmc04}.
There are also other models that incorporate quark degrees of freedom in
the study of nuclear many-body systems.
The quark mean-field (QMF) model~\cite{qmf98} takes the
constituent quark model for the nucleon instead of the MIT bag model, where
the constituent quarks interact with the meson fields created by other
nucleons. The QMF model has been successfully used for the description of
nuclear matter, finite nuclei, and hypernuclei~\cite{qmf00,qmf02,qmf05}.
Recently, the QMF model has been extended to a model based on
$SU(3)_{L}\times SU(3)_{R}$ symmetry and scale invariance~\cite{qmf0205}.
Using the Nambu-Jona-Lasinio model to describe the nucleon as a
quark-diquark state, it is also possible to discuss the stability of nuclear
matter based on the QMC idea~\cite{njl01}. The main advantage of these
models is their simplicity and self-consistency in taking into account quark
degrees of freedom in the study of nuclear many-body systems.

In this paper, we take the nontopological soliton bag model originally
proposed by Friedberg and Lee~\cite{fl77}, which is also called
Friedberg-Lee model in the literature, for the description of nucleons in
nuclear medium. In the Friedberg-Lee model, the nucleon is described as a
bound state of three quarks in a nontopological soliton formed by a scalar
field with nonlinear self-interactions. The Friedberg-Lee model has the
benefits that it is manifestly covariant and it exhibits dynamical bag
formation due to the coupling of quarks to the phenomenological scalar
field. Furthermore, it includes the MIT bag as a special case. The soliton
bag model has been extensively used to study the properties and structure of
hadrons in free space~\cite%
{fl82,fl83,Lubeck86,Lubeck87,fl86,fl90w,fl90,fl92,su05}. It has also been
applied to discuss the medium modification of nucleon properties~\cite%
{prc85,jpg93} and to study the dense matter properties within the
Wigner-Seitz approximation~\cite{npa00}. In the present work, we develop a
model to study the properties of nuclear matter and finite nuclei by
describing the nuclear many-body system as a collection of nontopological
soliton bags. The quarks inside the soliton bag couple not only to the
scalar field that binds the quarks together into nucleons, but also to
additional meson fields generated by the nuclear environment. The nucleons
interact through the self-consistent exchange of these mesons treated as
classical fields in the spirit of the QMC and QMF models. Because in this
model the soliton solution is significantly changed by the additional meson
fields, it is possible to investigate the modification of nucleon properties
in nuclear medium.

This paper is arranged as follows. In Sec.~\ref{sec:2}, we briefly describe
the Friedberg-Lee model for nucleons both in free space and in nuclear
medium, and then discuss the properties of nuclear matter as a collection of
nontopological soliton bags. In Sec.~\ref{sec:3}, we show the properties of
finite nuclei with the nucleons described as soliton bags, and also discuss
how the nucleon properties change inside nuclei. Section~\ref{sec:4} is
devoted to a summary.


\section{Nuclear matter as nontopological soliton bags}

\label{sec:2}

In this section, we first give a brief description of the Friedberg-Lee
model for individual nucleons. Then we develop a model for nuclear matter
regarded as a collection of nontopological soliton bags. This model enables
us to investigate possible modifications of nucleon properties in nuclear
medium.

\subsection{Free nucleon}
\label{sec:2a}

In the Friedberg-Lee model, a single nucleon is described as a bound state
of three quarks in a nontopological soliton formed by a phenomenological
scalar field with nonlinear self-interactions. The Friedberg-Lee model in
its simplest form is implemented through the effective Lagrangian density
\begin{equation}
\mathcal{L}=\bar{{\psi }}(i\gamma _{\mu }\partial ^{\mu }-m-g\phi ){\psi }+%
\frac{1}{2}\partial _{\mu }\phi \partial ^{\mu }\phi -U({\phi }) ,
\label{eq:lag0}
\end{equation}
where $\psi $ denotes the quark field. The quark mass $m$ is usually taken
to be zero for $u$ and $d$ quarks. $\phi$ is a color-singlet scalar field
that may be interpreted as the phenomenological representation of quantum
excitations of the self-interacting gluon field. The self-interaction of the
scalar soliton field is described by the potential
\begin{equation}
U(\phi )=\frac{a}{2!}{\phi }^{2}+\frac{b}{3!}{\phi }^{3}+\frac{c}{4!}{\phi }%
^{4}+B .  \label{eq:pot}
\end{equation}%
Here, the polynomial terminates in fourth order to ensure
renormalizability. The constants $a$, $b$, and $c$ are fixed within
a range so that $U({\phi })$ has a local minimum at $\phi =0$ and a
global minimum at $\phi =\phi_{v}$. The constant $B$ is determined
to make $U({\phi_{v}})=0$, and then the value $U(0)=B$ is to be
identified with the bag constant or volume energy density of a
cavity. We note that $\phi_{v}$ is the value of the soliton field in
the physical vacuum where the quark gets a mass $\sim g\phi_{v}$
from its coupling with the soliton field. Inside the nucleon where
valence quarks exist, the soliton field $\phi$ is reduced to be near
zero that means the perturbative vacuum is restored. It is
energetically favorable for three quarks to be localized in a cavity
in the soliton field, referred to as the soliton bag, so that the
nucleon appears as a bubble in the vacuum. In the mean-field
approximation, the soliton field is treated as a classical field
which is a time-independent \emph{c}-number field $\phi \left(
\mathbf{r}\right) $. The quark field operator is expanded in a
complete orthogonal set of Dirac spinor functions as
$\psi=\sum\limits_{k} b_k \psi_k $, where $b_k$ is the fermion
annihilation operator. For a nucleon, the three valence quarks are
in the lowest Dirac state $\psi_0$, then $\phi$ and $\psi_0$ satisfy
the coupled differential equations
\begin{equation}
\left(-i\vec{\alpha}\cdot \mathbf{\nabla}+g\beta \phi\right) \psi_{0}
=\epsilon_{0}\psi_{0} ,  \label{eq:free1}
\end{equation}
\begin{equation}
-\nabla^{2}\phi +\frac{\partial U(\phi)}{\partial \phi} =-3g\bar{\psi}%
_{0}\psi_{0} .  \label{eq:free2}
\end{equation}
The coupled equations have to be solved numerically. For the Dirac wave function
of the $1s_{1/2}$ quark, we use the notation%
\begin{equation}
\psi _{0}=\left(
\begin{array}{c}
u\left( r\right) \\
i\vec{\sigma}\cdot \widehat{\mathbf{r}} v\left( r\right)%
\end{array}%
\right) \chi ,  \label{eq:wf}
\end{equation}%
with $\chi =\left(
\begin{array}{c}
1 \\
0%
\end{array}%
\right) $ or $\left(
\begin{array}{c}
0 \\
1%
\end{array}%
\right) $ being the Pauli spinor. The total energy of the nucleon is given
by
\begin{equation}
E=3\epsilon_{0}+4\pi \int dr \ r^{2}\left[ \frac{1}{2}\left( \frac{d\phi }{dr%
}\right) ^{2}+U(\phi )\right] .  \label{eq:ne}
\end{equation}
The mean-square charge radius of the proton is given by%
\begin{equation}
\langle r^{2}\rangle =4\pi \int dr \ r^{4}\left( u^{2}+v^{2}\right) ,
\label{eq:nr}
\end{equation}%
and the proton magnetic moment is given by
\begin{equation}
\mu _{p}=\frac{8\pi }{3}\int dr \ r^{3}uv .  \label{eq:nmu}
\end{equation}%
The ratio of the axial-vector to vector coupling constants is given by
\begin{equation}
g_{A}/g_{V}=\frac{20\pi }{3}\int dr \ r^{2}\left( u^{2}-\frac{1}{3}%
v^{2}\right) .  \label{eq:ngagv}
\end{equation}

\subsection{Center-of-mass correction}
\label{sec:2b}

There are several methods for calculating the center-of-mass (c.m.)
correction to nucleon properties in the soliton bag
model~\cite{fl83,Lubeck86,Lubeck87}. In this paper, we consider two
approaches for the c.m. correction so that we can estimate how
sensitive the results are to the c.m. correction approach used.
First, we adopt an operationally simpler approach based on the
relativistic energy-momentum relation to take into account the c.m.
correction, which has been extensively discussed in
Ref.~\cite{fl83}. The rest mass of the nucleon in this approach is
given by
\begin{equation}
M=\sqrt{E^{2}-\langle \mathbf{P}^{2}\rangle },  \label{eq:nm1}
\end{equation}
where $\mathbf{P}$ is the total-momentum operator. The corrected
root-mean-squared (rms) radius is given by
\begin{equation}
r_{c}=\sqrt{\left[ 1-\frac{2\epsilon _{0}}{E}+\frac{3\epsilon _{0}^{2}}{E^{2}%
}\right] \langle r^{2}\rangle +\frac{3}{2E^{2}}}.  \label{eq:nrc}
\end{equation}
We also use the Peierls-Yoccoz projection technique to evaluate the
c.m. correction to nucleon properties. Following the method
described in Ref.~\cite{Lubeck86}, the rest mass of the nucleon is
given by the expectation value of the Hamiltonian in the
zero-momentum projected state. In this approach, the nucleon state
is assumed to be the direct product of a soliton coherent state and
a three-quark state
\begin{equation}
|N\rangle =e^{\lambda A^{\dagger }}b_{1}^{\dagger }b_{2}^{\dagger
}b_{3}^{\dagger }|0\rangle .  \label{eq:stn}
\end{equation}
The soliton coherent state is given by
\begin{equation}
e^{\lambda A^{\dagger }}|0\rangle =\prod_{\mathbf{k}}\exp \left[ \sqrt{\frac{%
\omega _{k}}{2}}f_{\mathbf{k}}a_{\mathbf{k}}^{\dagger }\right] |0\rangle ,
\label{eq:coh}
\end{equation}
where $a_{\mathbf{k}}^{\dagger }$ is the creation operator of the soliton
field. The soliton field operator can be expanded in terms of
$a_{\mathbf{k}}$ and $a_{\mathbf{k}}^{\dagger }$
\begin{equation}
\phi (\mathbf{r})=\phi _{v}+\left( 2\pi \right) ^{-3/2}\int d^{3}k\frac{1}{%
\sqrt{2\omega _{k}}}\left( a_{\mathbf{k}}e^{i\mathbf{k\cdot r}}+a_{\mathbf{k}%
}^{\dagger }e^{-i\mathbf{k\cdot r}}\right) .
\end{equation}
The expectation value of the soliton field in the coherent state has exactly
the properties of the mean-field value $\phi $ obtained by solving
Eqs.~(\ref{eq:free1}) and (\ref{eq:free2}). The soliton coherent state is
localized, so the nucleon state has no definite momentum. To construct a
momentum eigenstate, we use the Peierls-Yoccoz projection. The zero-momentum
projected state is given by
\begin{equation}
|\mathbf{P}=0\rangle =\int d^{3}X|\mathbf{X}\rangle ,  \label{eq:stp}
\end{equation}
where $|\mathbf{X}\rangle $ is a nucleon state localized at the point
$\mathbf{X}$
\begin{equation}
|\mathbf{X}\rangle =e^{\lambda A^{\dagger }(\mathbf{X})}b_{1}^{\dagger }(%
\mathbf{X})b_{2}^{\dagger }(\mathbf{X})b_{3}^{\dagger }(\mathbf{X})|0\rangle
.  \label{eq:stx}
\end{equation}%
The operator $A^{\dagger }(\mathbf{X})$ can be written as
\begin{equation}
\lambda A^{\dagger }(\mathbf{X})=\int d^{3}k\sqrt{\frac{\omega _{k}}{2}}f_{%
\mathbf{k}}(\mathbf{X})a_{\mathbf{k}}^{\dagger }.  \label{eq:adag}
\end{equation}%
For a translationally invariant operator $O$, its expectation value in the
zero-momentum projected state is given by
\begin{equation}
\langle O\rangle =\frac{\langle \mathbf{P}=0|O|\mathbf{P}=0\rangle }{\langle
\mathbf{P}=0|\mathbf{P}=0\rangle }=\frac{\int d^{3}Xd^{3}Y\langle \mathbf{X}%
|O|\mathbf{Y}\rangle }{\int d^{3}Xd^{3}Y\langle \mathbf{X}|\mathbf{Y}\rangle
}=\frac{\int d^{3}Z\langle -\frac{1}{2}\mathbf{Z}|O|\frac{1}{2}\mathbf{Z}%
\rangle }{\int d^{3}Z\langle -\frac{1}{2}\mathbf{Z}|\frac{1}{2}\mathbf{Z}%
\rangle },
\end{equation}%
where $\mathbf{Z}=\mathbf{Y}-\mathbf{X}$. The normalization condition can be
expressed as
\begin{equation}
\langle -\frac{1}{2}\mathbf{Z}|\frac{1}{2}\mathbf{Z}\rangle =N_{\phi }(%
\mathbf{Z})N_{q}(\mathbf{Z})^{3},  \label{eq:opz}
\end{equation}%
where
\begin{equation}
N_{\phi }(\mathbf{Z})=\exp \left[ \int d^{3}k\frac{\omega _{k}}{2}f_{\mathbf{%
k}}^{\ast }\left( -\frac{1}{2}\mathbf{Z}\right) f_{\mathbf{k}}\left( \frac{1%
}{2}\mathbf{Z}\right) \right] ,  \label{eq:onz}
\end{equation}%
\begin{equation}
N_{q}(\mathbf{Z})=\int d^{3}r\psi _{0}^{\dagger }\left( \mathbf{r+}\frac{1}{2%
}\mathbf{Z}\right) \psi _{0}\left( \mathbf{r-}\frac{1}{2}\mathbf{Z}\right) .
\label{eq:nqz}
\end{equation}%
Therefore, the nucleon mass is given by the expectation value of the
Hamiltonian in the zero-momentum projected state
\begin{equation}
M=\langle :H:\rangle =\langle :H_{q}+H_{q\phi }+H_{\phi }:\rangle .
\label{eq:nm2}
\end{equation}%
These expectation values of normal-ordered products are
\begin{equation}
\langle :H_{q}+H_{q\phi }:\rangle =3\int d^{3}ZN_{\phi }(\mathbf{Z})N_{q}(%
\mathbf{Z})^{2}\frac{\int d^{3}r\psi _{0}^{\dagger }\left( \mathbf{r}+\frac{1%
}{2}\mathbf{Z}\right) [-i\vec{\alpha }\cdot \mathbf{\nabla }+g\beta \bar{%
\phi}(\mathbf{r};\mathbf{Z})]\psi _{0}\left( \mathbf{r}-\frac{1}{2}\mathbf{Z}%
\right) }{\int d^{3}ZN_{\phi }(\mathbf{Z})N_{q}(\mathbf{Z})^{3}},
\end{equation}%
\begin{equation}
\langle :H_{\phi }:\rangle =\frac{\int d^{3}ZN_{\phi }(\mathbf{Z})N_{q}(%
\mathbf{Z})^{3}\mathcal{E}_{\phi }(\mathbf{Z})}{\int d^{3}ZN_{\phi }(\mathbf{%
Z})N_{q}(\mathbf{Z})^{3}},
\end{equation}%
where
\begin{equation}
\mathcal{E}_{\phi }(\mathbf{Z})=\int d^{3}r\left[ \frac{1}{2}\bar{\pi}(%
\mathbf{r};\mathbf{Z})^{2}+\frac{1}{2}|\mathbf{\nabla }\bar{\phi}(\mathbf{r};%
\mathbf{Z})|^{2}+U(\bar{\phi})\right] .
\end{equation}%
The expectation values of the soliton field operators are given by
\begin{equation}
\bar{\phi}(\mathbf{r};\mathbf{Z})=\frac{1}{2}\left[ \phi \left( \mathbf{r-}%
\frac{1}{2}\mathbf{Z}\right) +\phi \left( \mathbf{r+}\frac{1}{2}\mathbf{Z}%
\right) \right] ,
\end{equation}%
\begin{equation}
\bar{\pi}(\mathbf{r};\mathbf{Z})=-i\left( 2\pi \right) ^{-3/2}\int d^{3}k%
\frac{\omega _{k}}{2}\left[ f_{\mathbf{k}}\left( \frac{1}{2}\mathbf{Z}%
\right) e^{i\mathbf{k\cdot r}}-f_{\mathbf{k}}^{\ast }\left( -\frac{1}{2}%
\mathbf{Z}\right) e^{-i\mathbf{k\cdot r}}\right] .
\end{equation}%
We note that the corrected rms radius in this approach has the same
expression as given by Eq.~(\ref{eq:nrc})~\cite{Lubeck87}.

\subsection{Parameters in the Friedberg-Lee model}
\label{sec:2c}

\begin{figure}[htb]
\includegraphics[bb=40 170 540 630, width=8.6 cm, clip]{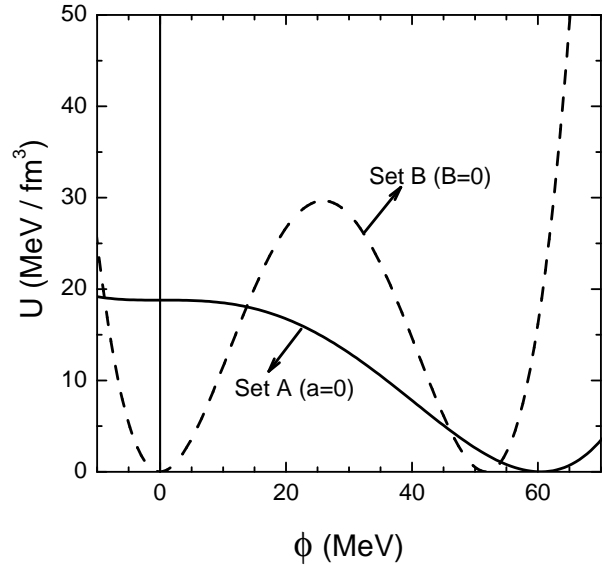}
\caption{The self-interaction of the soliton field as a function
of $\protect\phi$ for the two parameter sets characterized by $a=0$ and $B=0$.}
\label{fig:energy}
\end{figure}

In the Friedberg-Lee model, the parameters $a$, $b$, $c$, and $g$ are
constrained by reproducing reasonable nucleon properties. We adjust the
parameters to fit the nucleon mass $M=939$ MeV and rms radius $r_{c}=0.83$
fm. In the present work, we take two sets of parameters using the
c.m. correction given by Eq.~(\ref{eq:nm1}). Set A: $a=0$, $%
b=-79.61$ fm$^{-1}$, $c=780$, $g=13.7$ is characterized by $a=0$,
where $U({\phi })$ has a inflection point at $\phi =0$. Set B:
$a=69.945$ fm$^{-2}$, $b=-1600$ fm$^{-1}$, $c=12200$, $g=24.55$ is
characterized by $B=0$, where the relationship among the parameters
$b^{2}=3ac$ is obtained by requiring $B=0$. The parameters in the
Friedberg-Lee model have to be in a range where $U({\phi })$ has a
local minimum at $\phi =0$ and a global minimum at $\phi =\phi_{v}$.
Therefore, sets A and B correspond to the two limiting cases that
can be seen in Fig.~\ref{fig:energy}.

In order to compare the results with different c.m. correction
approaches, we take another parameter set using the projection
technique for the
c.m. correction given by Eq.~(\ref{eq:nm2}). Set C: $a=19.08$ fm$%
^{-2}$, $b=-335.02$ fm$^{-1}$, $c=1961.0$, $g=24.55$ is also
characterized by $B=0$. We use the same coupling constant $g$ in
sets B and C. By comparing the results with these two parameter
sets, we can estimate how sensitive the results are to the c.m.
correction approach used.

We compute nucleon properties using these three parameter sets. The
nucleon mass $M=939$ MeV and rms radius $r_{c}=0.83$ fm are obtained because
they are constrains for the parameter sets. Set A gives the proton magnetic
moment $\mu_{p}=2.80$ and the ratio of the axial-vector to vector coupling
constants $g_{A}/g_{V}=0.87$. Set B predicts $\mu _{p}=2.77$ and $%
g_{A}/g_{V}=0.90$, whereas set C gives $\mu_{p}=2.85$ and $g_{A}/g_{V}=0.80$.
We note that the experimental values are $\mu_{p}=2.79$ and $g_{A}/g_{V}=1.25$.
It is shown that these three parameter sets in the Friedberg-Lee model
can give reasonable results for nucleon properties in free space.

\subsection{Nuclear matter}
\label{sec:2d}

We develop a model for nuclear many-body system based on the
Friedberg-Lee model in the spirit of the QMC model. Nuclear matter in this
model is considered as a collection of nontopological soliton bags. The
solitons interact through the self-consistent exchange of $\sigma$, $\omega$,
and $\rho$ mesons that are treated as classical fields in the mean-field
approximation. For a soliton embedded in nuclear matter, the quarks inside
the nucleon couple not only to the soliton field $\phi$ which binds the
quarks together into the nucleon, but also to additional meson fields
$\sigma$, $\omega$, and $\rho$ generated by other nucleons in nuclear medium.
We assume that the meson mean fields $\sigma$, $\omega$, and $\rho $ can be
regarded as constants in uniform matter and the soliton field $\phi$
that serves to bind the quarks together does not participate in nucleon-nucleon
interactions. Therefore, $\phi$ depends on spatial coordinates inside a
nucleon, whereas $\sigma $, $\omega $, and $\rho $ are constants. With the
presence of these additional meson fields in nuclear matter, the quark and
soliton fields in the nucleon satisfy the coupled equations
\begin{equation}
\left( -i\vec{\alpha}\cdot \mathbf{\nabla}+g\beta \phi +g_{\sigma
}^{q}\beta \sigma +g_{\omega }^{q}\omega +g_{\rho }^{q}\tau _{3}\rho \right)
\psi _{0} =\widetilde{\epsilon}_{0}\psi _{0} ,  \label{eq:im1}
\end{equation}
\begin{equation}
-\nabla ^{2}\phi +\frac{\partial U(\phi )}{\partial \phi } =-3g\bar{\psi }%
_{0}\psi _{0} ,  \label{eq:im2}
\end{equation}%
where $g_{\sigma }^{q}$, $g_{\omega}^{q}$, and $g_{\rho}^{q}$ are the
coupling constants of the $\sigma$, $\omega$, and $\rho$ mesons with
quarks, respectively. We solve the coupled equations and calculate the
in-medium nucleon properties analogously to the case of free nucleons.
Here, the constant $\sigma$ field provides an additional scalar potential
to the quarks and as a consequence changes the solutions $\psi_{0}$ and $\phi$
from those obtained by Eqs.~(\ref{eq:free1}) and (\ref{eq:free2}). On the
other hand, the $\omega$ and $\rho$ fields do not cause any changes of
$\psi_{0}$ and $\phi$ except to shift the energy level by a constant vector
potential, $\widetilde{\epsilon}_{0}\left( \sigma ,\omega ,\rho \right)=
\epsilon_{0}\left(\sigma\right)+g_{\omega}^{q}\omega+g_{\rho}^{q}\tau_{3}\rho$.
Hence, $\psi_{0}$ and $\phi$ can be expressed as a function of the $\sigma$
mean field.

Analogously to the case of free nucleons, we use two approaches to
take into account the c.m. correction for nucleons in nuclear
matter. The first approach is based on the relativistic
energy-momentum relation~\cite{npa90}, in which the effective
nucleon mass is given by
\begin{equation}
M^{\ast}\left(\sigma\right)=\sqrt{E\left(\sigma\right)^{2}-\langle\mathbf{P}^{2}\rangle} ,
\label{eq:nms1}
\end{equation}
where
\begin{equation}
E\left(\sigma\right)=3\epsilon_{0}\left(\sigma\right)
                    +4\pi \int dr\text{ }r^{2}\left[ \frac{1}{2}
\left( \frac{d\phi}{dr}\right) ^{2}+U(\phi)\right] .
\label{eq:imne}
\end{equation}
The second approach is based on the Peierls-Yoccoz projection technique
as described in Sec.~\ref{sec:2b}. The effective nucleon mass is given by
the expectation value of the Hamiltonian in the zero-momentum projected state
\begin{equation}
M^{\ast }\left( \sigma \right) = \langle :H:\rangle
=\frac{\langle \mathbf{P}=0|:H:|\mathbf{P}=0\rangle }
{\langle\mathbf{P}=0|\mathbf{P}=0\rangle } ,
\label{eq:nms2}
\end{equation}
which can be calculated analogously to the case of free nucleons. We
note that the quark wave function $\psi_{0}$ and the soliton field
$\phi$ are altered by the $\sigma$ mean field in nuclear matter.
Therefore, the calculated nucleon properties are different from
those in free space. They can be expressed as functions of the
$\sigma$ mean field. The density dependence of these quantities is
obtained by a self-consistent determination of $\sigma$ at a given
nuclear matter density. The prescription of Eq.~(\ref{eq:nms1}) has
also been used in the QMC and QMF models for removing the c.m.
motion~\cite{qmc95,qmc97,qmf00}. Another prescription used in the
QMC model is to incorporate the c.m. correction into the parameter
$z_0$ and assume that $z_0$ is independent of the matter
density~\cite{qmccm1}.

\begin{figure}[htb]
\includegraphics[bb=30 30 570 730,width=8.6 cm, clip]{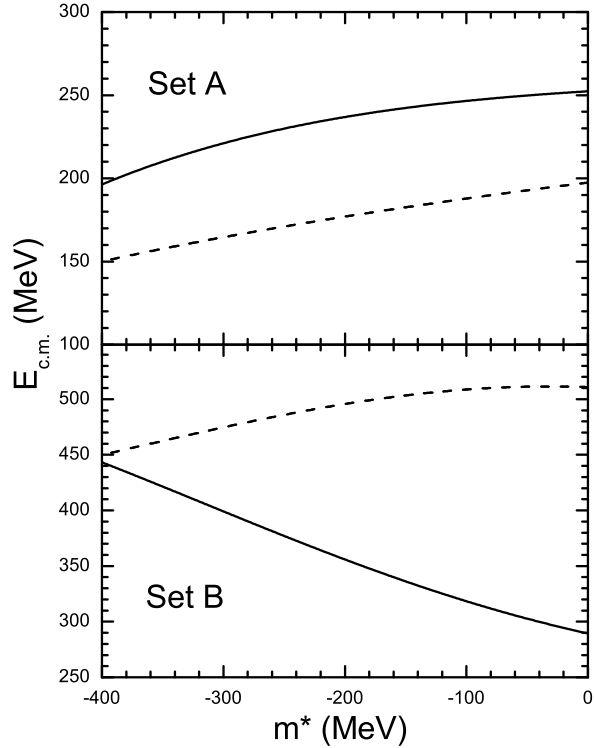}
\caption{The c.m. energy of the nucleon in nuclear matter,
$E_{\mathrm{c.m.}}=E-M^{\ast}$, as a function of the effective quark
mass, $m^{\ast}=g_{\sigma}^{q}\sigma$. The solid lines show the
results of the first c.m. correction approach where $M^{\ast}$ is
given by Eq.~(\ref{eq:nms1}). The dashed lines show the results of
the second c.m. correction approach with $M^{\ast}$ given by
Eq.~(\ref{eq:nms2}). We note that $m^{\ast}=0$ in free space and
$m^{\ast}\sim -240$ MeV at normal nuclear matter density.}
\label{fig:cm}
\end{figure}

It is interesting to compare the two c.m. correction approaches used
in this paper and investigate how the c.m. correction changes in
nuclear matter. We define the c.m. energy of a nucleon in nuclear
matter as $E_{\mathrm{c.m.}}=E-M^{\ast}$, where $M^{\ast}$ is given
by Eq.~(\ref{eq:nms1}) in the first c.m. correction approach and by
Eq.~(\ref{eq:nms2}) in the second one. In Fig.~\ref{fig:cm}, we show
$E_{\mathrm{c.m.}}$ as a function of the effective quark mass
$m^{\ast}=g_{\sigma}^{q}\sigma$, which is proportional to the
$\sigma$ mean field. The results of the first c.m. correction
approach are shown by the solid lines, while those of the second
approach are shown by the dashed lines. We have $m^{\ast}=0$ in free
space and $m^{\ast}\sim -240$ MeV at normal nuclear matter density.
It is seen that the variation of $E_{\mathrm{c.m.}}$ in nuclear
matter depends on the c.m. correction approach and the parameter
set. As the density increases, $E_{\mathrm{c.m.}}$ obtained in the
first c.m. correction approach (solid lines) decreases in the upper
panel and increases in the lower panel, whereas the results of the
second approach (dashed lines) slowly decreases in both panels of
Fig.~\ref{fig:cm}. We find that the tendency of $E_{\mathrm{c.m.}}$
with the second c.m. correction approach is quite similar to that
shown in Fig.~A.1 of Ref.~\cite{qmccm1} which is the results of a
serious calculation for a relativistic harmonic oscillator
potential. Therefore, the treatment for the c.m. correction by the
projection technique is considered to be more reliable than the
simple c.m. correction based on the relativistic energy-momentum
relation.

To investigate in-medium nucleon properties and nuclear matter
characteristics, we take a hybrid treatment for nuclear matter.
The effective nucleon mass and couplings are obtained at the quark level,
whereas the nucleon Fermi motion is treated at the hadron level.
To perform a many-body calculation for nuclear matter, we start from the
effective Lagrangian at the hadron level within the mean-field approximation
\begin{eqnarray}
\mathcal{L}_{\mathrm{RMF}} &=&\bar{\psi}\left[ i\gamma _{\mu }\partial ^{\mu
}-M^{\ast }\left( \sigma \right) -g_{\omega }\gamma ^{0}\omega -g_{\rho
}\gamma ^{0}\tau _{3}\rho \right] \psi  \nonumber \\
&&-\frac{1}{2}m_{\sigma }^{2}\sigma ^{2}+\frac{1}{2}m_{\omega }^{2}\omega
^{2}+\frac{1}{2}m_{\rho }^{2}\rho ^{2}\text{ },  \label{eq:ml}
\end{eqnarray}%
where $\psi$ denotes the nucleon field. The effective nucleon mass
$M^{\ast}\left( \sigma \right) $ is obtained in the Friedberg-Lee
model, which has been given by Eq.~(\ref{eq:nms1}) in the first c.m.
correction approach and by Eq.~(\ref{eq:nms2}) in the second one.
The nonvanishing meson fields are replaced by their expectation
values $\sigma =\left\langle \sigma \right\rangle$, $\omega
=\left\langle \omega^{0}\right\rangle$, $\rho =\left\langle
\rho^{03}\right\rangle$ that are constants in a static infinite
nuclear matter. The nucleon-meson couplings are related to the
quark-meson couplings as $g_{\omega }=3g_{\omega }^{q}$ and $g_{\rho
}=g_{\rho }^{q}$~\cite{qmf00}. From the Lagrangian given by
Eq.~(\ref{eq:ml}), we obtain the equations of motion for nucleons
and mesons in nuclear matter,
\begin{equation}
\left[ i\gamma _{\mu }\partial ^{\mu } \right. - \left. M^{\ast }\left(
\sigma \right) -g_{\omega }\gamma ^{0}\omega -g_{\rho }\gamma ^{0}\tau
_{3}\rho \right] \psi = 0 ,  \label{eq:dirac0}
\end{equation}
\begin{equation}
m_{\sigma }^{2}\sigma = -\frac{\partial M^{\ast }\left( \sigma \right) }{%
\partial \sigma }\langle \bar{\psi}\psi \rangle ,  \label{eq:kgs0}
\end{equation}
\begin{equation}
m_{\omega }^{2}\omega = g_{\omega }\langle \bar{\psi}\gamma ^{0}\psi \rangle
,  \label{eq:kgw0}
\end{equation}
\begin{equation}
m_{\rho }^{2}\rho = g_{\rho }\langle \bar{\psi}\gamma ^{0}\tau _{3}\psi
\rangle .
\label{eq:kgr0}
\end{equation}
With $M^{\ast}\left( \sigma \right)$ obtained at the quark level,
we solve the coupled equations self-consistently, and then
calculate the nuclear matter properties and medium modifications of nucleon
properties. In the present model, the quark-meson couplings $g_{\sigma }^{q}$,
$g_{\omega}^{q}$, and $g_{\rho }^{q}$ are determined by reproducing the nuclear
matter equilibrium density ($0.15$ fm$^{-3}$), energy per nucleon ($-16$ MeV),
and symmetry energy ($35$ MeV). The meson masses are taken to be
$m_{\sigma}=500$ MeV, $m_{\omega }=783$ MeV, and $m_{\rho }=770$ MeV.
We list in Table~\ref{tab:matter} the resulting nuclear matter properties
corresponding to the parameter sets A, B, and C used in the Friedberg-Lee
model. It is shown that the present model can provide a satisfactory
description of nuclear matter properties.

\begin{figure}[htb]
\includegraphics[ width=10 cm, clip]{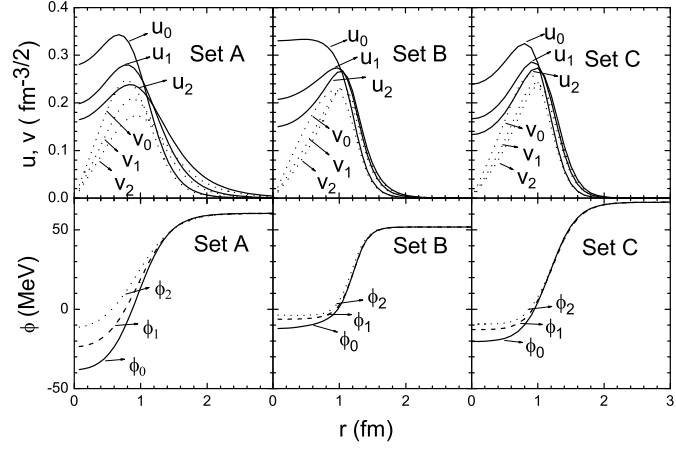}
\caption{The quark wave functions ($u$, $v$) and the scalar field $\protect\phi$
as a function of the nucleon radius ($r$) with the three parameter sets. The
results of the nucleon in free space are denoted by $u_0$, $v_0$ and $%
\protect\phi_0$. The results at the density $\protect\rho_0$ and $2\protect%
\rho_0$ are denoted by $u_1$, $v_1$, $\protect\phi_1$ and $u_2$, $v_2$, $%
\protect\phi_2$, respectively, where $\protect\rho_0=0.15$ fm$^{-3}$
is the normal nuclear matter density.} \label{fig:wave}
\end{figure}

Having determined the $\sigma$ mean field self-consistently in
nuclear matter, we investigate in-medium nucleon properties with the
quark wave function obtained by solving Eqs.~(\ref{eq:im1}) and
(\ref{eq:im2}). In Fig.~\ref{fig:wave} we plot the Dirac solutions
($u$, $v$) and the soliton field ($\phi$) as functions of the
nucleon radius ($r$) for the three parameter sets used. The results
of the nucleon in free space are denoted by $u_0$, $v_0$, $\phi_0$,
while those at the density $\rho_0$ and $2\rho_0$ are denoted by
$u_1$, $v_1$, $\phi_1$ and $u_2 $, $v_2$, $\phi_2$, respectively.
Here $\rho_0=0.15$ fm$^{-3}$ is the normal nuclear matter density.
By comparing the results in medium with those in free space, we find
that the quark wave functions in nuclear matter are significantly
different from those in free space. This leads to the medium
modifications of nucleon properties because they are calculated
using the quark wave functions. In the present model, the quark wave
function depends on the $\sigma$ mean field, hence the density
dependence of nucleon properties is obtained through the density
dependence of the $\sigma$ mean field. It is interesting to compare
the results obtained with the three parameter sets used. Set A gives
stronger in-medium modifications than sets B and C, especially at
larger $r$. The results of set B is quite similar to those of set C
because both of them are characterized by $B=0$. It is seen in
Fig.~\ref{fig:wave} that the soliton field ($\phi$) around the
surface at high density in the case of set A is much broader than
that in set B. This yields more leakage of the quark wave function
in set A than in set B. It may lead to the difference in the density
dependence of nucleon properties, such as $r_c$ and $g_{A}/g_{V}$,
because they are sensitive to the behavior of the quark wave
function at large $r$. By comparing with the sharp boundary of the
MIT bag model, we find that set B is closer to the MIT bag model
than set A. Therefore, the results of set B are rather close to
those of the MIT bag model. This indicates that the surface behavior
of the confinement potential plays an important role in determining
nucleon properties.

\begin{figure}[htb]
\includegraphics[bb=20 20 540 470, width=8.6 cm, clip]{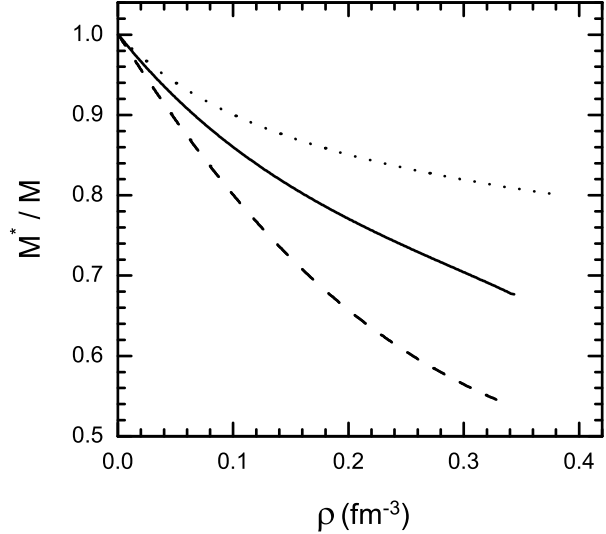}
\caption{The ratio of the effective nucleon mass in nuclear matter to that
in free space, $M^{\ast } / M$, as a function of nuclear matter density $%
\protect\rho $. The results with the parameter sets A, B, and C are shown
by the solid, dashed, and dotted lines, respectively.}
\label{fig:mass}
\end{figure}

In Fig.~\ref{fig:mass}, we present the ratio of the effective
nucleon mass in nuclear matter to that in free space, $M^{\ast }/M$,
as a function of nuclear matter density $\rho$. The solid and dashed
lines correspond to the cases of set A and set B where the effective
nucleon mass $M^{\ast}$ is given by Eq.~(\ref{eq:nms1}) using the
first c.m. correction approach. The dotted line corresponds to the
case of set C where $M^{\ast}$ is given by Eq.~(\ref{eq:nms2}) using
the second c.m. correction approach. We note that the effective
nucleon mass in the present model is calculated at the quark level,
which is not a simple linear function of $\sigma$ as given in the
Walecka model~\cite{WS86}. It is more like the characteristics of
the QMC model. As shown in Fig.~\ref{fig:mass}, the effective
nucleon mass decreases with increasing density, and the results
depend on the parameter set used. The large difference between sets
B and C is mainly due to the different c.m. correction approaches
used in these two cases. The first c.m. correction approach used in
the case of set B provides a strong increase of $E_{\mathrm{c.m.}}$
in nuclear matter (see solid line in the lower panel of
Fig.~\ref{fig:cm}), whereas the second c.m. correction approach used
in the case of set C gives a weak decrease of $E_{\mathrm{c.m.}}$
(see dashed line in Fig.~\ref{fig:cm}). This leads to the difference
in $M^{\ast}$ between sets B and C due to the relation
$M^{\ast}=E-E_{\mathrm{c.m.}}$. Therefore, the drop of $M^{\ast}$ in
set B is much more rapid than that in set C. On the other hand, the
difference between sets A and B can be understood as a consequence
of the different behaviors of $E_{\mathrm{c.m.}}$ in these two cases
(compare solid lines in the upper and lower panels of
Fig.~\ref{fig:cm}), though the first c.m. correction approach is
used in both sets A and B.

\begin{figure}[htb]
\includegraphics[bb=20 20 540 470, width=8.6 cm, clip]{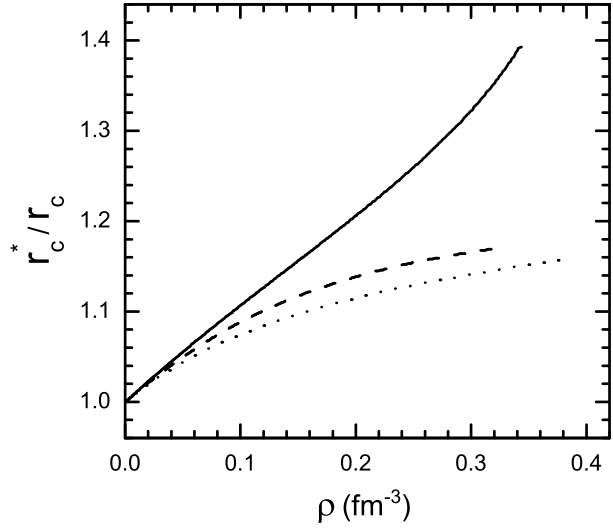}
\caption{The ratio of the nucleon rms radius in nuclear matter to that in
free space, $r_c^{\ast } / r_c$, as a function of nuclear matter density $%
\protect\rho$. The lines are labeled as in Fig.~\protect\ref{fig:mass}.}
\label{fig:r}
\end{figure}

\begin{figure}[htb]
\includegraphics[bb=20 20 540 470, width=8.6 cm, clip]{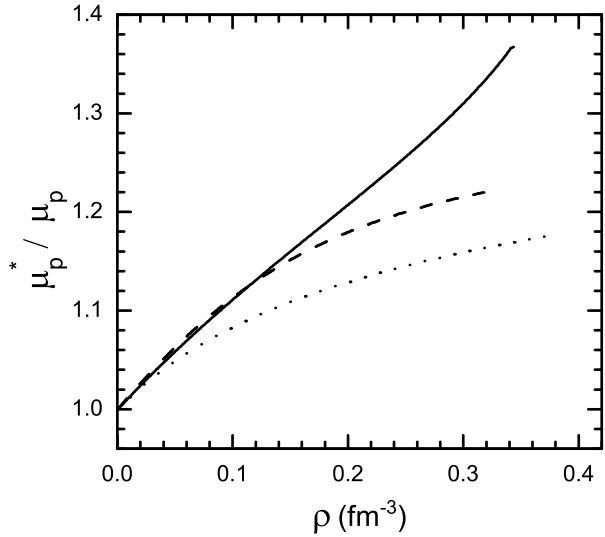}
\caption{The ratio of the proton magnetic moment in nuclear matter to that
in free space, $\protect\mu_{p}^{\ast} / \protect\mu_{p}$, as a function of
nuclear matter density $\protect\rho$. The lines are labeled as in Fig.~%
\protect\ref{fig:mass}.}
\label{fig:mu}
\end{figure}

\begin{figure}[htb]
\includegraphics[bb=20 20 540 470, width=8.6 cm, clip]{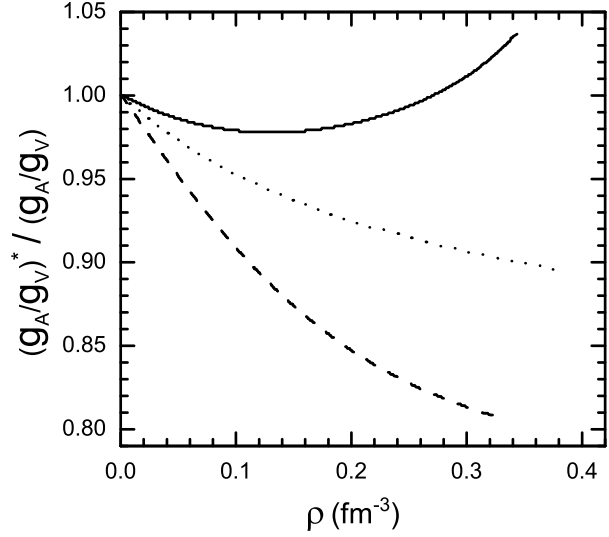}
\caption{The ratio of the axial-vector to vector coupling constants in
nuclear matter $(g_{A}/g_{V})^\ast$ to that in free space $g_{A}/g_{V}$ as a
function of nuclear matter density $\protect\rho$. The lines are labeled as
in Fig.~\protect\ref{fig:mass}.}
\label{fig:gagv}
\end{figure}

We show in Fig.~\ref{fig:r} the ratio of the nucleon rms radius in
nuclear matter to that in free space, $r_c^{\ast} / r_c$, as a
function of nuclear matter density $\rho$. It is very interesting to
see the expansion of the nucleon size in medium. We find the nucleon
rms radius increases by about $10$-$16\%$ at normal nuclear matter
density. This result is quite different from those obtained in other
models. For example, the QMC model predicts only $1$-$3\%$
enhancement in the nucleon rms radius at normal nuclear matter
density~\cite{qmc96}, and the QMF model gives about $5$-$9\%$
increase~\cite{qmf00}. The chiral quark-soliton model predicts a
$2.4\%$ enhancement, while the swelling constrained by quasielastic
inclusive electron-nucleus scattering data is less than
$6\%$~\cite{cqs04}. We note that although the Friedberg-Lee model
can give similar quark distributions and nucleon properties to those
obtained in the MIT bag model in free space, the quarks satisfy
different equations and boundary conditions in these two models, and
therefore the swelling of the nucleon rms radius could be quite
different between the QMC model and the present calculation. In
Fig.~\ref{fig:mu}, we present the ratio of the proton magnetic
moment in nuclear matter to that in free space, $\mu_{p}^{\ast} /
\mu_{p}$, as a function of nuclear matter density $\rho$. It is
shown that the results depend on the parameter set used. The
difference between different parameter sets increases with
increasing density. Set A gives nearly the same low-density behavior
as set B, but rather different results at high density. The
enhancement in set C is smaller than those in sets A and B. We show
in Fig.~\ref{fig:gagv} the ratio of the axial-vector to vector
coupling constants in nuclear matter $(g_{A}/g_{V})^\ast$ to that in
free space $g_{A}/g_{V}$ as a function of nuclear matter density
$\rho$. The results of set A decrease slightly at lower densities,
and then increase at higher densities. On the other hand, the
results of set B drop significantly with increasing density. The
results of set C is between sets A and B. At normal nuclear matter
density, we obtain $(g_{A}/g_{V})^\ast / (g_{A}/g_{V}) \simeq 0.98 $
for set A, $(g_{A}/g_{V})^\ast / (g_{A}/g_{V}) \simeq 0.87 $ for set
B, and $(g_{A}/g_{V})^\ast / (g_{A}/g_{V}) \simeq 0.94 $ for set C.
It is obvious that medium modifications of nucleon properties in
this work depend on the parameter set used in the Friedberg-Lee
model. Because set A gives stronger in-medium modifications of the
quark wave functions than sets B and C as shown in
Fig.~\ref{fig:wave}, we obtain larger increases in $r_c^{\ast}$ and
$\mu_{p}^{\ast}$ with set A, as shown in Figs.~\ref{fig:r} and
\ref{fig:mu}. On the other hand, the difference in
$(g_{A}/g_{V})^\ast$ shown in Fig.~\ref{fig:gagv} could be due to
the competition between changes of $u$ and $v$ according to
Eq.~(\ref{eq:ngagv}).


\section{Properties of finite nuclei}

\label{sec:3}

In this section, we extend the present model to study the properties of
finite nuclei and the modification of nucleon properties in a nucleus. The
nucleus is described as a collection of nontopological soliton bags that
interact through the self-consistent exchange of $\sigma$, $\omega$,
and $\rho$ mesons. In principle, these meson mean fields are functions of the
spatial coordinates in the nucleus, but it is rather complicated if the
variation of these quantities over the small nucleon volume is taken into
account. Therefore, we take some suitably averaged form for the meson mean
fields in order to make the numerical solution feasible.
We use the local density approximation which replace the meson mean fields
by their value at the center of the nucleon and neglect the spatial variation
of the mean fields over the small nucleon volume~\cite{qmc96,qmc98,qmf00}.
The equations of motion for nucleons
and mesons in a spherically symmetric nucleus are given by
\begin{equation}
\left[ i\gamma _{\mu }\partial ^{\mu }-M^{\ast }\left( \sigma \right)
-g_{\omega }\gamma ^{0}\omega -g_{\rho }\gamma ^{0}\tau _{3}\rho -e\frac{%
(1+\tau _{3})}{2}\gamma ^{0}A\right] \psi =0 ,  \label{eq:dirac1}
\end{equation}
\begin{equation}
\left( -\Delta +m_{\sigma }^{2}\right) \sigma =-\frac{\partial M^{\ast
}\left( \sigma \right) }{\partial \sigma }\langle \bar{\psi}\psi \rangle ,
\label{eq:kgs1}
\end{equation}
\begin{equation}
\left( -\Delta +m_{\omega }^{2}\right) \omega =g_{\omega }\langle \bar{\psi}%
\gamma ^{0}\psi \rangle ,  \label{eq:kgw1}
\end{equation}
\begin{equation}
\left( -\Delta +m_{\rho }^{2}\right) \rho =g_{\rho }\langle \bar{\psi}\gamma
^{0}\tau _{3}\psi \rangle ,  \label{eq:kgr1}
\end{equation}
\begin{equation}
-\Delta A =e\langle \bar{\psi}\frac{(1+\tau _{3})}{2}\gamma ^{0}\psi \rangle
,  \label{eq:kga1}
\end{equation}%
where the mean fields are functions of the radial coordinate of the nucleon
center in the nucleus. We solve the preceding equations self-consistently
with the effective nucleon mass obtained at the quark level.

\begin{figure}[htb]
\includegraphics[bb=20 20 540 470, width=8.6 cm, clip]{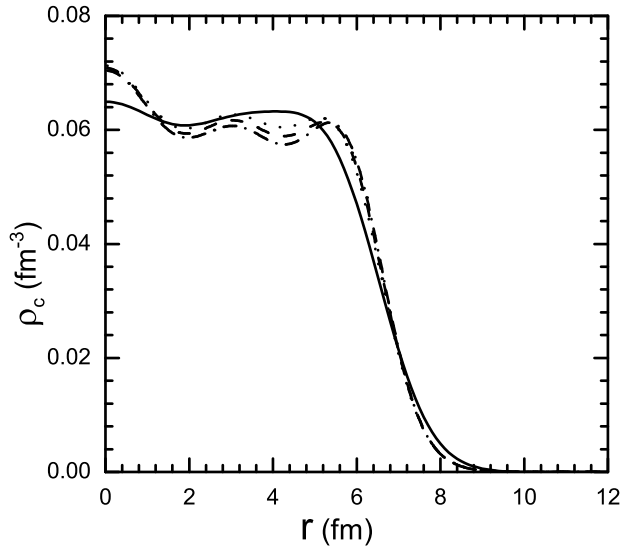}
\caption{The charge density distributions for $^{208}\text{Pb}$ compared
with the experimental data (solid line)~\protect\cite{EXP}.
The results with the parameter sets A, B, and C are shown by the dashed,
dotted, and dot-dashed lines, respectively.}
\label{fig:charge208}
\end{figure}

We present the numerical results of several spherical nuclei. In
Table~\ref{tab:nuclei}, the calculated binding energies per nucleon
and rms charge radii are compared with the experimental
values~\cite{ST94}. By comparing the results of sets A, B, and C, we
find that the binding energies of set C are larger than those of
sets A and B. This is mainly because set C gives smaller
incompressibility ($K=184$ MeV) than sets A ($K=302$ MeV) and B
($K=384$ MeV) when they have the same saturation density
($\rho_0=0.15$ fm$^{-3}$) and energy per particle ($E/A=-16$ MeV) as
shown in Table~\ref{tab:matter}. We list the calculated
single-particle energies and spin-orbit splittings in
Tables~\ref{tab:spectrum} and~\ref{tab:splitting}, and compare with
the experimental data taken from Ref.~\cite{EXPSO}. It is seen that
all calculated spin-orbit splittings are smaller than the
experimental data, while set C gives the smallest values among the
three parameter sets. This is due to the large effective nucleon
masses obtained in the present model as shown in
Table~\ref{tab:matter}. According to analysis with many
quantum-hadrodynamics models, there exists a tight correlation
between the spin-orbit splitting of finite nuclei and the effective
nucleon mass in nuclear matter at saturation density, and the
spin-orbit splitting increases with decreasing $M^{\ast}$. It is
known that $M^{\ast}/M \sim 0.6$ is required in order to reproduce
the empirical spin-orbit splittings of finite nuclei. This is the
reason why we get small spin-orbit splittings in the present model
and the smallest values in set C. This shortage might be improved by
including nonlinear meson self-interactions as taken in the QMC
model~\cite{qmc98}. We plot in Fig.~\ref{fig:charge208} the
resulting charge density distributions for $^{208}$Pb and compare
with the experimental values~\cite{EXP}. As seen in this figure, the
calculated results are in good agreement with the experimental
values, and there is no explicit difference between the three
parameter sets.

\begin{figure}[htb]
\includegraphics[bb=20 20 540 470, width=8.6 cm, clip]{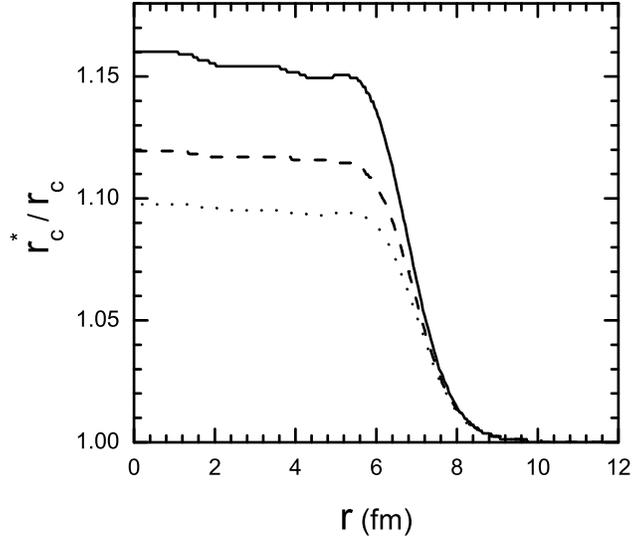}
\caption{The ratio of the proton rms radius in $^{208}\text{Pb}$ to that in
free space as a function of radial coordinate $r$.
The results with the parameter sets A, B, and C are shown
by the solid, dashed, and dotted lines, respectively.}
\label{fig:rms208}
\end{figure}

\begin{figure}[htb]
\includegraphics[bb=20 20 540 470, width=8.6 cm, clip]{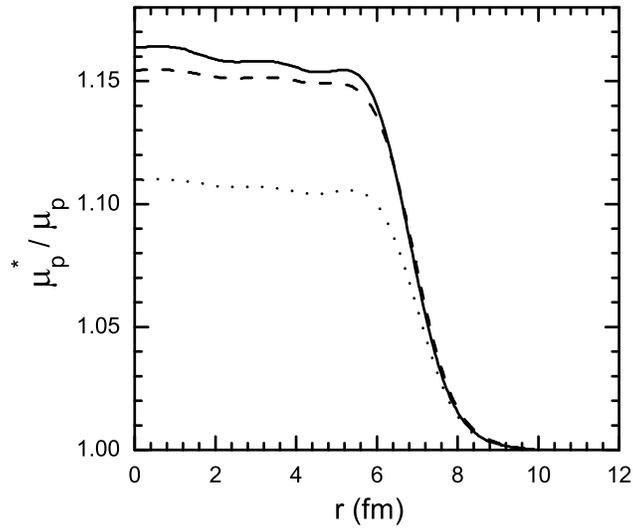}
\caption{The ratio of the proton magnetic moment in $^{208}\text{Pb}$ to
that in free space as a function of radial coordinate $r$. The lines are
labeled as in Fig.~\protect\ref{fig:rms208}.}
\label{fig:mu208}
\end{figure}

It is also possible to investigate the modification of nucleon properties in
finite nuclei. Using the local density approximation, the nucleon
properties at the radial coordinate $r$ in a nucleus are obtained through
the values of $\sigma \left( r\right)$, because the nucleon properties in
this model are functions of the $\sigma$ mean field obtained at the quark level.
In Figs.~\ref{fig:rms208} and \ref{fig:mu208}, we show the ratios of the
proton rms radius and magnetic moment in $^{208}$Pb to those in free space
as functions of the radius $r$. It is found that the proton radius and magnetic
moment increase significantly at the center of $^{208}$Pb. These quantities
decrease to the values in free space from the center to the surface of the
nucleus. It is shown that the results depend on the parameter set used,
which is consistent with the results of nuclear matter shown in
Figs.~\ref{fig:r} and \ref{fig:mu}.


\section{Conclusion}

\label{sec:4}

By treating the nucleons as nontopological soliton bags, we have proposed a
model for the description of nuclear matter and finite nuclei, and at the
same time for the study of medium modifications of nucleon properties. The
nontopological soliton bag model, proposed by Friedberg and Lee, exhibits a
dynamical bag formation due to the coupling of quarks to the
phenomenological scalar field $\phi$. The quarks inside the soliton bag
couple not only to the scalar field $\phi$ that binds the quarks together
into nucleons, but also to additional meson fields $\sigma$, $\omega$, and
$\rho$ generated by the nuclear environment. The nucleons interact through
the self-consistent exchange of $\sigma$, $\omega$, and $\rho$ meson
fields that are treated as classical fields in the mean-field
approximation. This model enables us to investigate the medium modification
of nucleon properties because the soliton bag is significantly influenced by
the additional meson fields in nuclear medium.

We have considered two approaches for the c.m. correction to nucleon
properties so that we can estimate how sensitive the results are to
the c.m. correction approach used. We have adopted three parameter
sets in the Friedberg-Lee model that are constrained by reproducing
free nucleon properties. The quark-meson coupling constants are
fitted to reproduce the empirical saturation properties of nuclear
matter. The present model can provide a reasonable description of
nuclear matter. We have found that the properties of the nucleon are
significantly modified in nuclear medium. At normal nuclear matter
density, the nucleon radius increases by about $10$-$16\%$, while
the proton magnetic moment increases by about $11$-$16\%$. We have
applied the present model to study the properties of spherical
nuclei and found that it could give a reasonable description of the
ground state properties of finite nuclei. It is very interesting to
see the nucleon properties change from the nuclear surface to the
nuclear interior. The present model incorporates explicit quark
degrees of freedom into nuclear many-body systems. It is notable
that the quark structure of the nucleon plays a crucial role in the
description of nuclear matter and finite nuclei.


\section*{Acknowledgments}

This work was supported in part by the National Natural Science Foundation
of China (No. 10675064).


\newpage

\begin{table}[tbp]
\caption{The nuclear matter properties in the present model with the
parameter sets A, B, and C. The saturation density and the energy per
particle are denoted by $\protect\rho_0$ and $E/A$, the symmetry energy
by $a_{\mathrm{sym}}$, the incompressibility by $K$, and the effective
mass by $M^{\ast}$.}
\label{tab:matter}\vspace{0.0cm}
\par
\begin{center}
\begin{tabular}{cccccc}
\hline\hline
& $\rho_0$ & $E/A $ & $a_{\mathrm{sym}}$ & $K$ & $M^{\ast }/M$ \vspace{-0.2cm%
} \\
& (fm$^{-3})$ & (MeV) & (MeV) & (MeV) &  \\ \hline
Set A & 0.15 & -16.0 & 35 & 302 & 0.81 \\
Set B & 0.15 & -16.0 & 35 & 384 & 0.72 \\
Set C & 0.15 & -16.0 & 35 & 184 & 0.87 \\ \hline\hline
\end{tabular}%
\end{center}
\end{table}

\begin{table}[tbp]
\caption{The binding energy per nucleon $E/A$ and the rms charge radius $R_c$
for $^{40}\text{Ca}$, $^{90}\text{Zr}$, and $^{208}\text{Pb}$.}
\label{tab:nuclei}\vspace{0.0cm}
\par
\begin{center}
\begin{tabular}{cccccccccc}
\hline\hline
& \multicolumn{4}{c}{$E/A $ (\textrm{{MeV})}} &  & \multicolumn{4}{c}{$R_c$ (%
\textrm{{fm})}} \\ \cline{2-5}\cline{7-10}
\raisebox{2ex}[0ex] & Set A & Set B & Set C & Expt. &  & Set A & Set B & Set
C & Expt. \\ \hline
$^{40}\text{Ca} $ & 8.53 & 7.66 & 9.35 & 8.55 &  & 3.42 & 3.46 & 3.38 & 3.45
\\
$^{90}\text{Zr} $ & 8.36 & 7.81 & 8.97 & 8.71 &  & 4.26 & 4.27 & 4.24 & 4.26
\\
$^{208}\text{Pb}$ & 7.48 & 7.13 & 7.95 & 7.87 &  & 5.52 & 5.50 & 5.54 & 5.50
\\ \hline\hline
\end{tabular}%
\end{center}
\end{table}

\begin{table}[tbp]
\caption{The single-particle energies of proton ($p$) and neutron ($n$) for $%
^{40}\text{Ca}$. The experimental data are taken
from Ref.~\protect\cite{EXPSO}. All energies are in MeV.}
\label{tab:spectrum}\vspace{0.0cm}
\par
\begin{center}
\begin{tabular}{cccccccccccc}
\hline\hline
& \multicolumn{2}{c}{Set A} &  & \multicolumn{2}{c}{Set B} &  &
\multicolumn{2}{c}{Set C} &  & \multicolumn{2}{c}{Expt.} \\
\cline{2-3}\cline{5-6}\cline{8-9}\cline{11-12}
Shell & $p$ & $n$ &  & $p$ & $n$ &  & $p$ & $n$ &  & $p$ & $n$ \\ \hline
$1s_{1/2}$ & 35.6 & 43.7 &  & 37.8 & 45.9 &  & 34.7 & 42.8 &  & 50$\pm$11 &
50.0 \\
$1p_{3/2}$ & 24.6 & 32.4 &  & 25.7 & 33.5 &  & 24.4 & 32.2 &  & 34$\pm$6 &
30.0 \\
$1p_{1/2}$ & 23.5 & 31.3 &  & 23.6 & 31.4 &  & 23.7 & 31.6 &  & 34$\pm$6 &
27.0 \\
$1d_{5/2}$ & 12.5 & 20.0 &  & 12.8 & 20.3 &  & 12.6 & 20.2 &  & 15.5 & 21.9
\\
$1d_{3/2}$ & 10.4 & 18.0 &  & 9.3 & 16.8 &  & 11.4 & 19.0 &  & 8.3 & 15.6 \\
$2s_{1/2}$ & 8.1 & 15.6 &  & 7.9 & 15.2 &  & 8.4 & 15.9 &  & 10.9 & 18.2 \\
\hline\hline
\end{tabular}%
\end{center}
\end{table}

\begin{table}[tbp]
\caption{The spin-orbit splittings of proton ($\Delta E_p$) and neutron ($%
\Delta E_n$) for $^{40}\text{Ca}$ and $^{208}\text{Pb}$. The experimental
data are taken from Ref.~\protect\cite{EXPSO}. All quantities are in MeV.}
\label{tab:splitting}\vspace{0.0cm}
\par
\begin{center}
\begin{tabular}{lccccc}
\hline\hline
& \multicolumn{2}{c}{$^{40}\text{Ca}$} &  & \multicolumn{2}{c}{$^{208}\text{%
Pb}$} \\ \cline{2-3}\cline{5-6}
& $\Delta E_p$ & $\Delta E_n$ &  & $\Delta E_p$ & $\Delta E_n$ \\
& $(1d_{5/2}-1d_{3/2})$ & $(1d_{5/2}-1d_{3/2})$ &  & $(1g_{9/2}-1g_{7/2})$ &
$(2f_{7/2}-2f_{5/2})$ \\ \hline
Set A & 2.0 & 2.0 &  & 1.2 & 0.7 \\
Set B & 3.5 & 3.5 &  & 2.2 & 1.3 \\
Set C & 1.1 & 1.1 &  & 0.6 & 0.4 \\
Expt. & 7.2 & 6.3 &  & 4.0 & 1.8 \\ \hline\hline
\end{tabular}%
\end{center}
\end{table}


\end{document}